\documentclass[preprint]{elsarticle}

\usepackage{amssymb}
\usepackage{subfigure}
\usepackage{lineno}
\journal{Nuclear Instruments and Methods A}

\begin{document}
\begin{frontmatter}
\title{Studies of VERITAS Photomultipliers After Eight Years of Use}
\author{D. Hanna, S. O'Brien, T. Rosin}
\address{Department of Physics,
McGill University,
Montreal, QC H3A2T8, Canada} 
 
\begin{abstract}

The VERITAS gamma-ray telescope array has been operating since 2007 and has 
been equipped with Hamamatsu R10560-100-20 PMTs since 2012. 
A decision to continue operations into the mid 2020s was taken in 2019 so the
question of whether the PMTs would need replacing became important and a study
was initiated.

We present results from scanning two groups of 20 Hamamatsu R10560-100-20  
PMTs with an LED flasher.
One group comprised five PMTs from each of the four VERITAS telescopes 
and the other was made up of
20 PMTs of the same type, and date of manufacture, that had never been used. 
We measured three test variables related to gains and high-voltage response and 
found that there were no significant differences between the two groups.
This indicates that there
has been little ageing in the PMTs that have been used on the telescopes and 
that replacement is unnecessary.

\end{abstract}
\end{frontmatter}

%\linenumbers

\section{Introduction}

The Very Energetic Radiation Imaging Telescope Array System 
(VERITAS)~\cite{v_ref} is a ground-based detector for gamma-ray astronomy.
It comprises four large imaging atmospheric Cherenkov telescopes (IACTs) 
deployed at the Fred Lawrence Whipple Observatory, approximately 65 km south 
of Tucson, AZ.
Each of the telescopes is based on a 12-metre-diameter Davies-Cotton
reflector that 
collects Cherenkov light from showers of relativistic particles initiated by 
astrophysical gamma rays when they impact the upper atmosphere.

The light is directed onto a `camera' in the form of a close-packed hexagonal
array of 499 photomultiplier tubes (PMTs), each 26 mm in diameter. 
The resulting pulses are used for triggering and are digitized and
saved for off-line analysis.
The pattern of light on the PMTs, acting as pixels, allows one to reconstruct
an image of the shower.
The position of the shower image and the size of the summed pulses are used 
to reconstruct the energy and arrival direction of the incident gamma ray.
The morphology of the shower image plays a crucial role in rejecting backgrounds
from showers caused by charged cosmic rays. 

The VERITAS array was commissioned in 2007 and was upgraded in 
2012~\cite{kieda, rajotte}.  
As part of the upgrade, all of the original 
Photonis XP2970 PMTs in the four cameras 
were replaced with Hamamatsu R10560-100-20 PMTs.
These have super-bialkali photocathodes, a new development at the time, with 
peak quantum efficiency near 35\%, a significant increase over the approximately
20\% of the Photonis PMTs.
As a result, the performance of VERITAS was improved, with a 
lower energy threshold and improved energy and angular resolution among the
benefits.

As part of the monitoring of the VERITAS instrument, the PMTs' gains are
measured on every observing night using a flasher 
system~\cite{flasher} based on 375 nm LEDs.
The flasher delivers light pulses to the camera with each PMT receiving the same
light level thanks to a diffuser on the flasher. 
Different light levels are obtained by stepping through the number of LEDs 
activated; the original flashers had 7 LEDs and were replaced in 2016 with 
15-LED flashers. 

Two gain-related parameters can be obtained from the flasher data.
The first is the `relative gain', which is a measure of the pulse size 
in a channel due to a given light level from the flasher.
The second is the `photostatistics gain' which is derived from the variances in 
pulse size as a function of light level. 
For details, see~\cite{flasher} and ~\cite{bencheikh}.

The photostatistics gain is the 
gain of the system starting at the first dynode of the 
PMT and ending after the digitizing electronics. 
It has units of picocoulomb per photoelectron.
The relative gain is similar but also includes the effect of the photocathode.
The flashers do not have monitors for estimating the absolute light levels
so it is not possible to make in-situ measurements of the photocathode
quantum efficiencies.
Differences between the two gain types can arise from photocathode differences;
a weak photocathode can be compensated for by increasing the amplification coming from the dynodes. 
According to our definitions, this would result in 
a higher photostatistics gain.

\begin{figure}
\centerline{\includegraphics[width=0.8\textwidth]{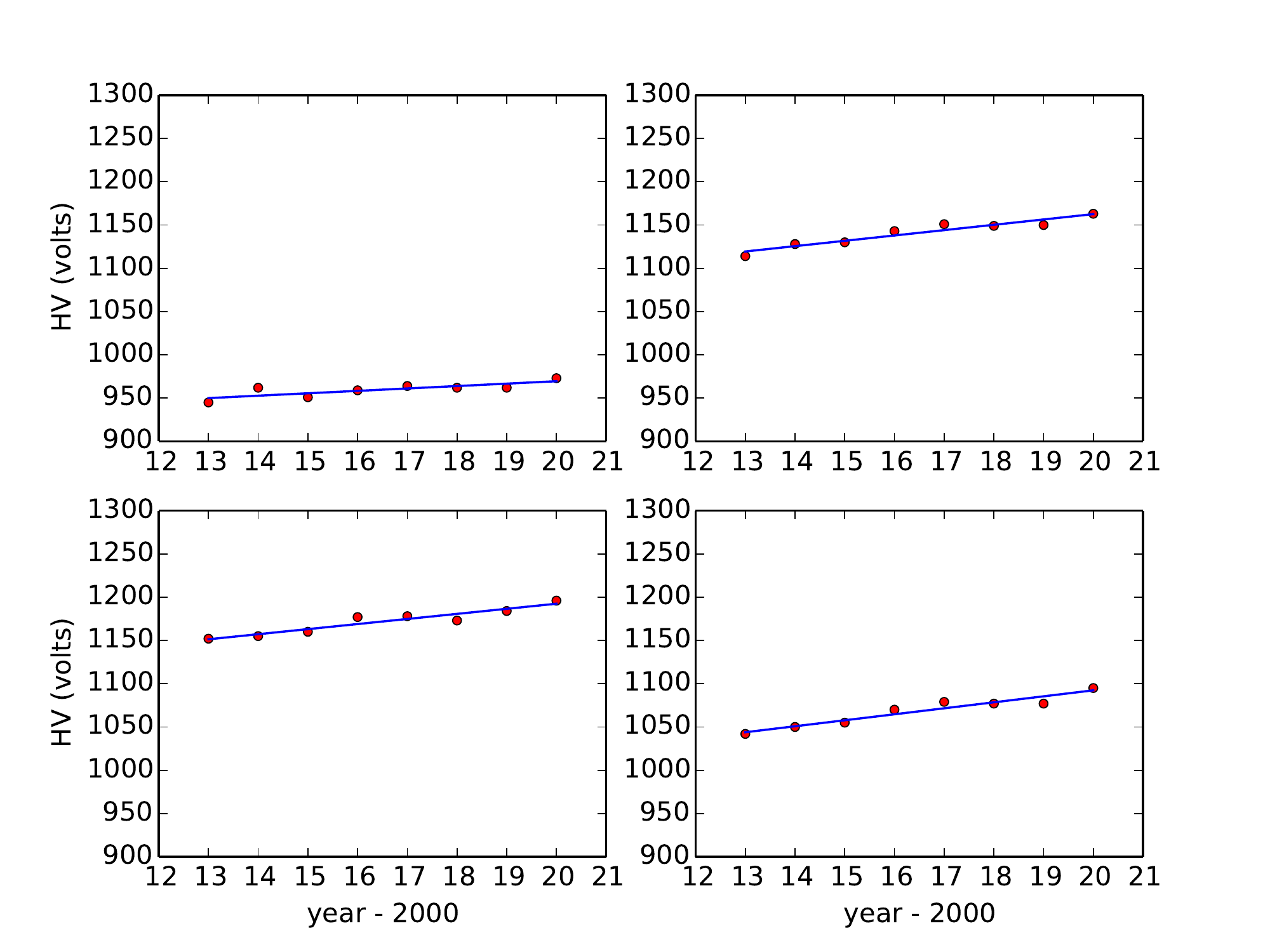}}
\vspace*{0.0cm}
\caption{Evolution of HV values for four typical PMTs from one of the VERITAS 
telescopes over the years using the values from November each year, starting in 
2013.
Each HV value results from making the PMT's 
relative gain equal to that of all the
other PMTs in the camera and from making the camera-averaged photostatics gain
equal to a constant value assumed in gamma-ray data analysis.
The HV values follow an approximately linear path, as shown by the fit.
}
\label{hva3}
\end{figure}

\begin{figure}
\centerline{\includegraphics[width=0.6\textwidth]{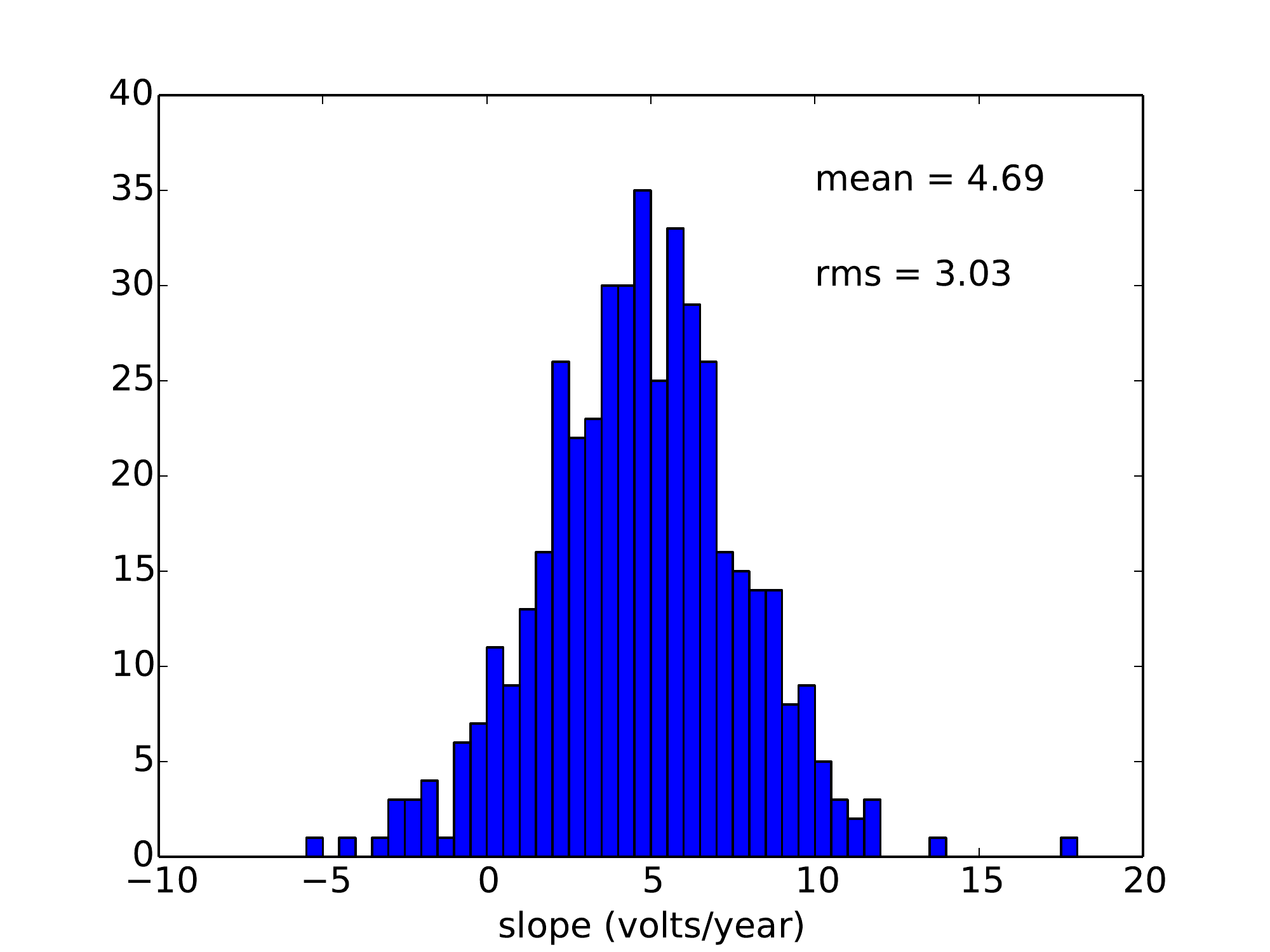}}
\vspace*{0.0cm}
\caption{A histogram of the slopes of fits like those shown in Fig.~\ref{hva3} for all PMTs in 
a single telescope. The other three VERITAS telescopes exhibit similar 
behaviour. 
}
\label{hva35}
\end{figure}

At the beginning of every observing season (typically from September through 
the following June), PMT high-voltages (HVs) are adjusted so that every PMT produces
the same signal level; this is called `flat-fielding' and it 
equalizes the relative gains of the PMTs.
The average photostatistics gain of the camera is also set, 
by adjusting the average level of the HVs.
Flat-fielding ensures a uniform camera response, necessary for triggering
purposes, while setting the photostatistics 
gain of the camera is needed for maintaining
agreement with the energy-scale used in the Monte-Carlo calculations that are
part of the data analysis chain.
This procedure has been used on the Hamamatsu PMTs 
for almost a decade and the long-term effects
can be seen by plotting the HVs vs time, as shown in Fig.~\ref{hva3},
where the HVs from an arbitrary flasher run in early November of each year 
are plotted vs year number for a sample of typical PMTs.
Slopes for all PMTs in the camera are shown in Fig.~\ref{hva35}.
The average slope is approximately 4.7 volts per year but the distribution is 
wide (RMS $\simeq$ 3.0 V/y).
The small number of slopes that are negative
gives an indication of the level of systematic effects in this 
diagnostic. 

It is important to note that the changes ones sees can be quite
significant over the long term.
A slope of 4.7 V/y means that the HV supplied to a PMT would need to increase 
by 47 V over a ten-year span, about 5\% for a PMT operating with HV near 
1000 V. 
Given the steep dependence of gain on HV
(shown later in this article to follow a power-law behaviour with an exponent
of approximately 5.7) this is equivalent to a 30\% change in gain.

\section{Laboratory Checks on PMTs}

\begin{figure}
\centerline{\includegraphics[width=0.6\textwidth]{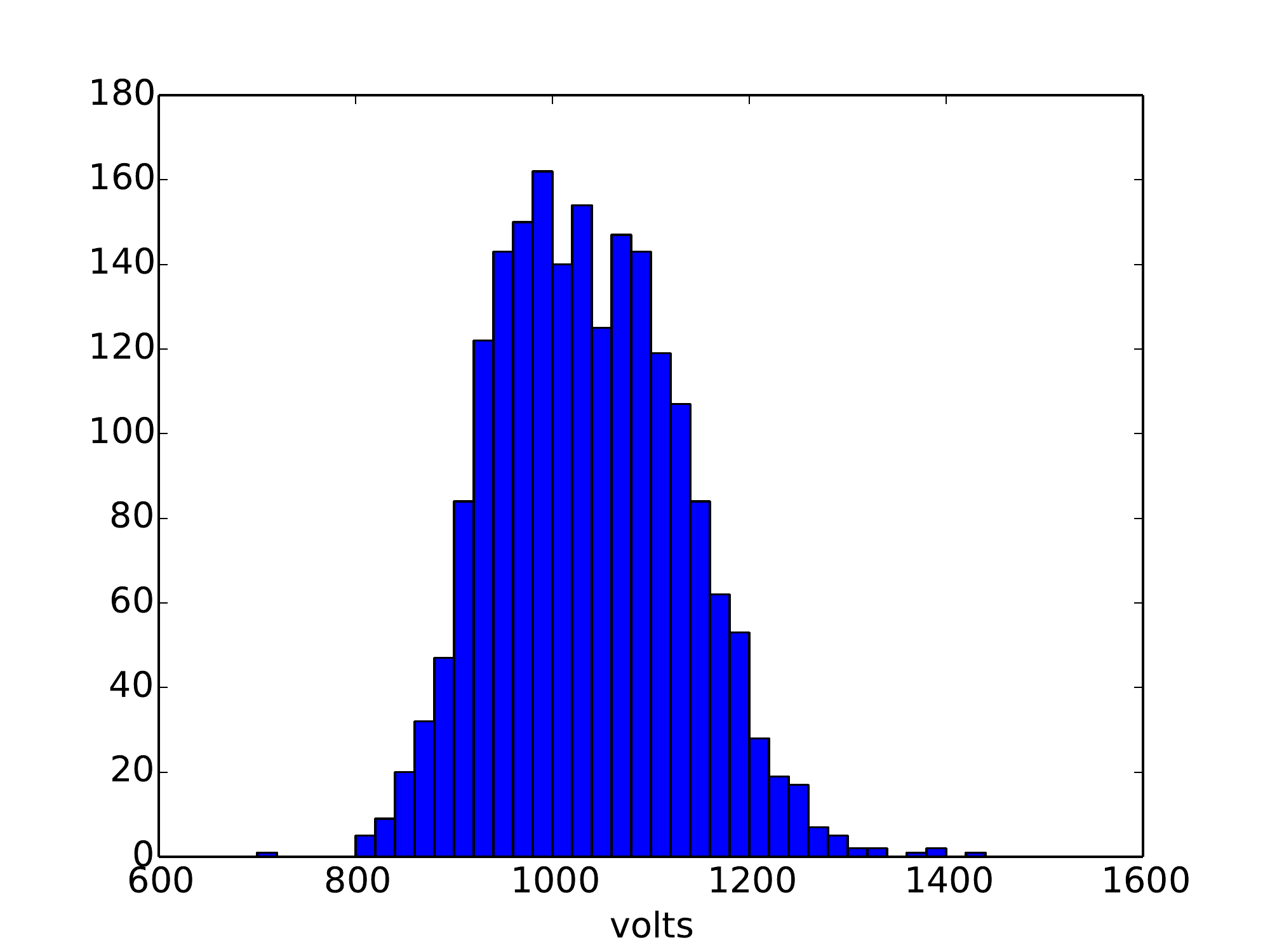}}
\vspace*{0.0cm}
\caption{
A histogram of the high-voltage values for all PMTs in all the VERITAS 
cameras, illustrating the wide range of values.
Yearly adjustments to the values, for flat-fielding and gain boosting are 
small compared to the width of this distribution and do not take the operating
voltages for any PMTs significantly outside of the range of these values.
}
\label{hva2}
\end{figure}

The need to increase, on average, the HV settings for the PMTs in order to 
maintain nominal gain values is a matter of interest, but not concern. 
It is important to keep the size of the increases in perspective, given that 
they amount to a few percent of the initial HV values and changes are much 
less than the spread of HV values over the whole set of PMTs.
This can be seen in Fig.~\ref{hva2} where the HV values for all PMTs in the
four telescopes are histogrammed.

It is natural to expect ageing effects in the VERITAS PMTs. 
The instrument is run for 1000 hours during a typical observing season and 
currents in the PMTs arising from night-sky-background are on the order of 
10 $\mu$A. 
The total charge collected on the last dynode of a PMT 
is therefore expected to be of order 10 - 100 C per season. 
It is therefore 
likely that damage to the last dynodes in the PMT is the reason for the
needed HV increases.
Indeed, this effect was studied before the R10560-100-20 PMTs were installed, with a 
similar PMT, the R9800-100-20~\cite{otte-icrc}. 
It was found that the gain dropped by about 25\%, from an initial value of 
$2 \times 10^5$, during a 700-hour test where a low intensity light source 
illuminated the PMT.
The initial current was 44 $\mu$A and the accumulated anode charge was 
approximately 96 C.
This result is not inconsistent with the gain-change effects discussed in the 
introduction.

However there could also be ageing effects at the photocathodes.
This has been checked twice since the PMTs were installed, once on a
sample of three PMTs from each telescope, done in 2014~\cite{finley}
and once on a sample of five PMTs from each telescope, comparing them to 20 
spare PMTs of the same type that had never been used~\cite{otte}, done in 2016.
The 2014 study found no evidence of damage to the photocathodes; indeed 
a fine-grained scan of the photocathode with an optical fibre revealed
no detectable changes in the topography of the photocathode.
The 2016 study concentrated more on the wavelength dependence of the 
photon detection efficiency;  
measurements were made at a single 10 mm-diameter spot at the centre of 
the PMT.
The conclusion of that work was that there was no significant performance 
degradation after four years of operation.

It was decided at the end of the 2019-20 observing season to check the PMTs
once again, their time-in-use having doubled since the 2016 study. 
This paper reports on the methods used for that check and on the results 
obtained. 

\section{Instrument Design}

The measurements made for this study were made during the month of July, 2020,
between the first and second 
wave of the Covid-19 Pandemic, as experienced in North America. 
During this time the same 20 PMTs as used in the 2016 study were extracted from
the VERITAS telescope cameras and were shipped to McGill University.
The 20 unused PMTs used in 2016 were also sent.
They were all measured using a new scanning setup constructed for fast 
parallel processing and limited operator intervention, motivated by laboratory
access rules in force at the time.

The basic idea was to illuminate the tested PMTs with a controlled light source 
while also illuminating a separate monitor PMT with the same source.
The monitor was used to control for any changes in the intensity of the light 
source over the course of the measurements and to serve as a standard to 
which the PMTs under test could be compared.

For the light source we used a 7-LED VERITAS flasher~\cite{flasher} 
and for the monitor PMT we used a spare Hamamatsu R10560-100-20 PMT.
To allow for faster processing, we built a setup that could be used to 
measure six PMTs 
(plus the monitor) at a time.

\subsection{Mechanics}

\begin{figure}[h]
\begin{tabular}{cc}
\begin{minipage}{0.45\linewidth}
\centerline{\includegraphics[width=1.0\linewidth]{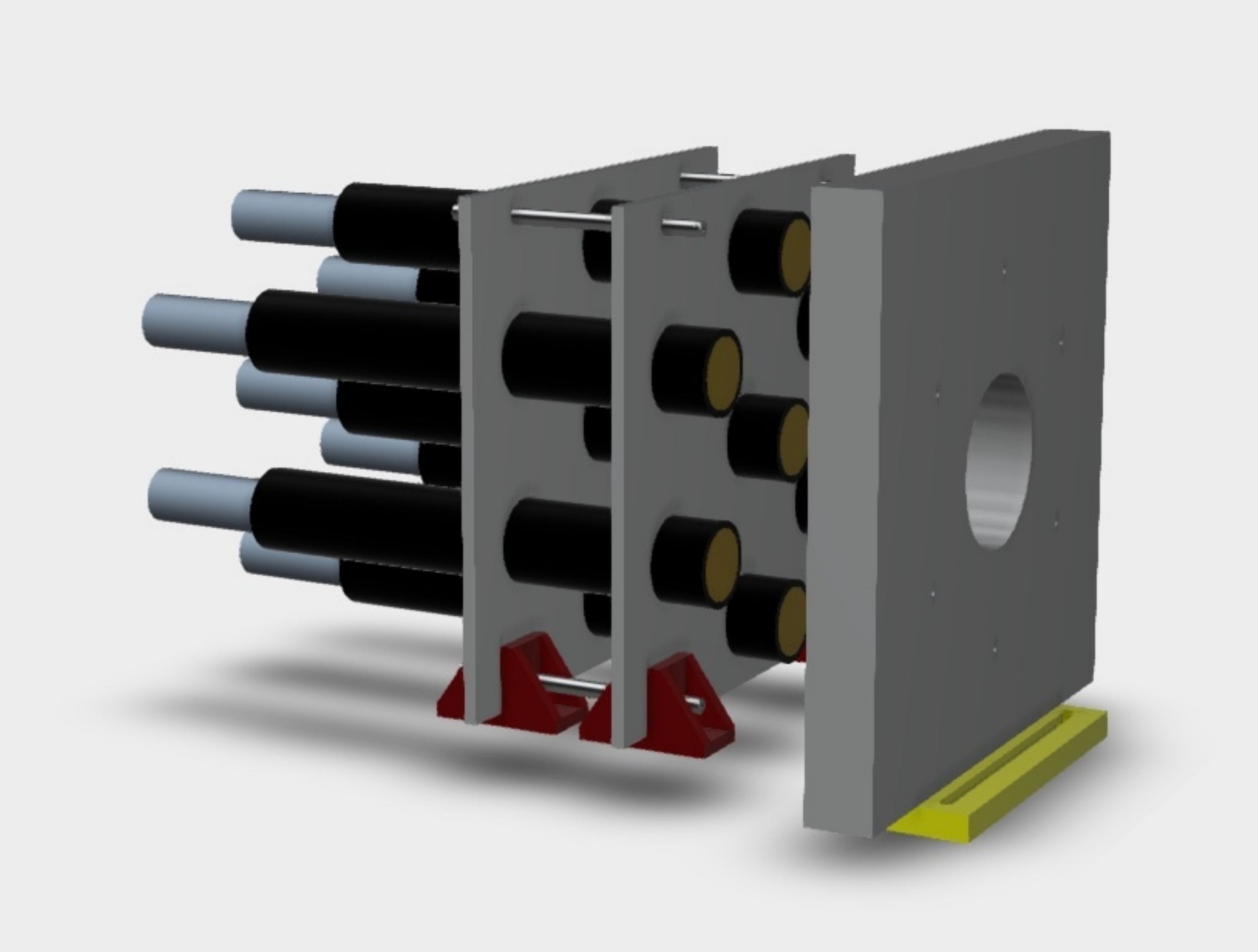}}
\vspace{0mm}
\end{minipage}

&

\begin{minipage}{0.45\linewidth}
\centerline{\includegraphics[width=0.91\linewidth]{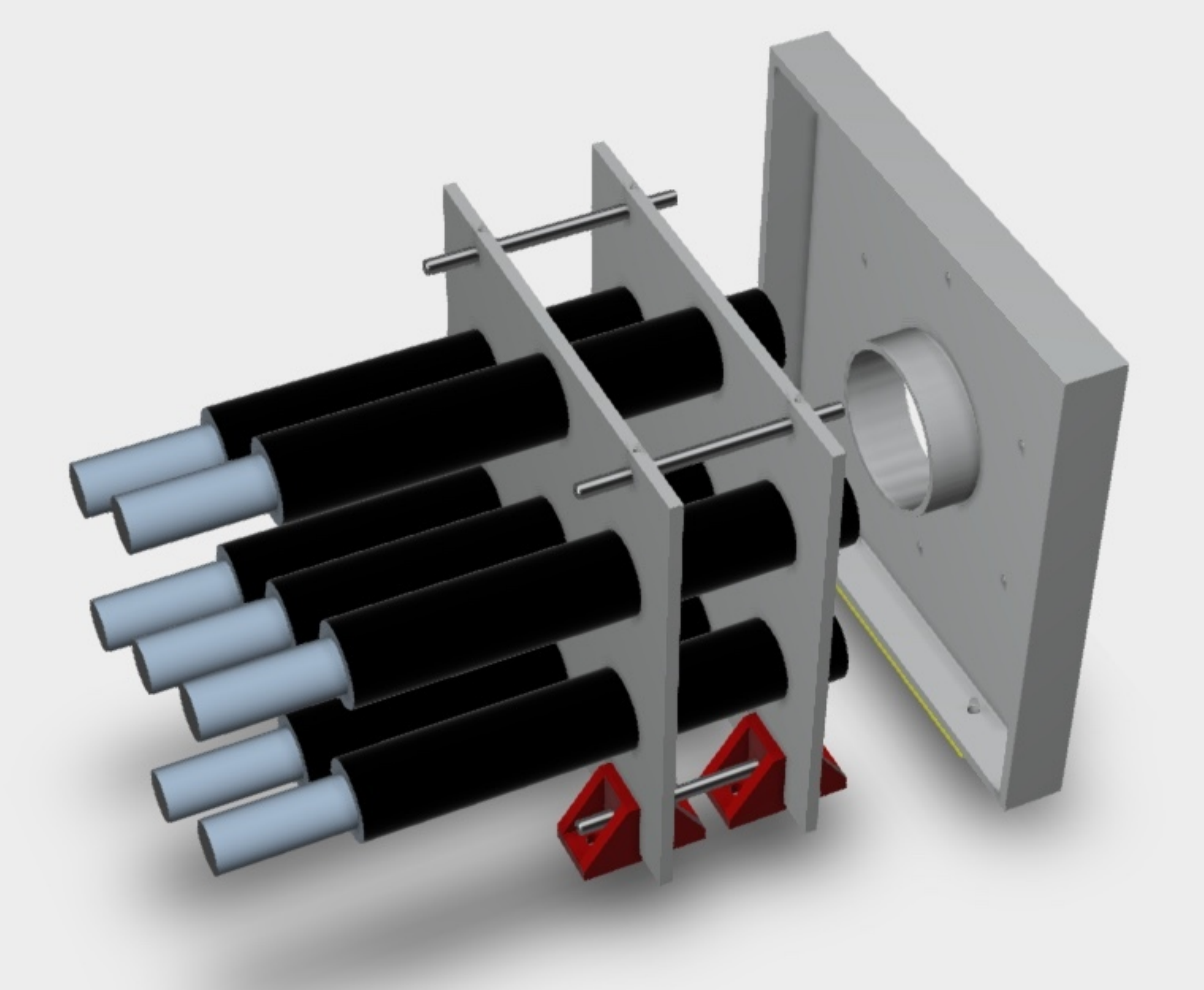}}
\vspace{0mm}
\end{minipage}
\end{tabular}
\caption{
Two renderings of the PMT test setup. The six PMTs under test are arranged 
in a hexagon pattern around the monitor PMT, which is at the centre.
A movable opaque 
screen with a 3-mm hole for each test PMT and a larger hole for the
monitor PMT is positioned immediately in front of the assembly. The screen is
shown here displaced longitudinally by 
a small distance for illustrative purposes.
In the right-hand view one can see a circular baffle that shields the test PMTs 
from scattered light.
}
\label{scanner}
\end{figure}

A rendering of the hardware is displayed in Fig.~\ref{scanner}.
The six PMTs under test were arranged in a hexagon around the monitor PMT 
at the centre, with the axis of each tested PMT  64 mm from the axis of 
the monitor.
An opaque 
screen that could be scanned in x and y was positioned immediately in front 
of the assembly such that a hole, 3 mm in diameter, lined up in front of each
test PMT.
Each hole controlled the region of illumination of its corresponding PMT and the
region was determined by the lateral position of the screen. 
A hole, 64 mm in diameter, in the centre of the screen allowed 
full illumination of the monitor
PMT, independent of the screen's lateral position. 
A baffle around this hole screened test PMTs from scattered light.
The screen was mounted on translation stages that could 
position it to a desired x, y coordinate using stepper motors.
The PMT assembly remained stationary.
All custom-built components were made from plastic on a 3D printer and the 
translation devices were commercially-available assemblies from Optics-Focus.
A single linear stage with 100 mm range was used for moving the screen 
horizontally and a stacked pair of motorized lab jacks, each with 50 mm range,
was used for moving the screen vertically. 
The cost of the scanner was approximately \$US 1400, dominated by the 
translation devices.

The flasher was positioned approximately 1 m away in the z direction. 
It comprised seven UV LEDs ($\lambda \simeq$ 375 nm) located behind an 
opal-glass diffuser.
The flasher was designed for deployment approximately 6 m away from the PMTs
in the VERITAS cameras so was too bright for the monitor PMT in this 
application. 
To deal with this we ran the monitor with reduced high-voltage and deployed 
a neutral-density filter of optical depth 1.5 in front of it.
The test PMTs were not overwhelmed by the flasher light thanks to the 
3-mm holes. 
With this arrangement, about 60 photoelectrons per pulse per LED were produced 
in each PMT.

The full face of the monitor PMT was exposed at all times so that any 
lateral variations in efficiency were averaged over, independent of the screen
position. 
Despite the neutral-density filter, the monitor received approximately 
16 times
more photons than the test PMTs so monitor pulse-height fluctuations due to 
photostatistics were not a concern.

The setup was housed in a light-tight box approximately 120 cm long by 60 cm 
wide by 60 cm high.

\subsection{Electronics and Data Acquisition}

The master clock for the scanning operations was provided by a Berkeley 
Nucleonics BNC-8010 pulse generator running at a rate of approximately 
420 Hz.
It provided a NIM-level pulse to trigger the flasher, which ramps through eight 
levels repetitively, incrementing with each
trigger pulse the number of LEDs that are powered.
The clock also triggered the readout of 7 Acqiris DC270 8-bit 1 GS/s digitizers
used to record the pulses from the PMTs.
The PMTs were supplied with high-voltage from two 4-channel CAEN N1419 
power supplies.

\subsection{Scanning Protocol}

The scanning procedure was run from a laptop computer that communicated with 
the CAEN HV modules, to set voltages, and the Acqiris digitizer system, that 
used an in-crate cPCI-6620 series Adlink computer.
The Adlink computer was used, via an Arduino single-board
microcontroller, to control 
the stepper motors that positioned the screen at each scan point.
Custom programs for controlling the digitizers and HV modules were written 
in Python. 

All scans proceeded in the same way, beginning with installation of six test
PMTs into the scanner.
The screen was positioned such that the 3-mm holes were aligned with the centres
of the PMTs. 
A rough value for HV for each PMT was determined using an oscilloscope and the 
PMTs were powered up. 
After a 20-minute warmup period, we set the high-voltages for the
PMTs in each batch by acquiring data at three HV settings and using a fit
to determine values that would lead to equal responses.
For the fit we used a function based on a simple model of the PMT 
that gives rise to the formula
$G/G_0 = (HV/HV_0)^\alpha$,
where $G$ ($G_0$) is the gain value at high-voltage value $HV$ ($HV_0$)
and $\alpha$ is a constant that is related to the number of 
dynodes~\cite{hamamatsu}. 

\begin{figure}[h!]
\vspace{0cm}
\centerline{\includegraphics[width=0.7\linewidth]{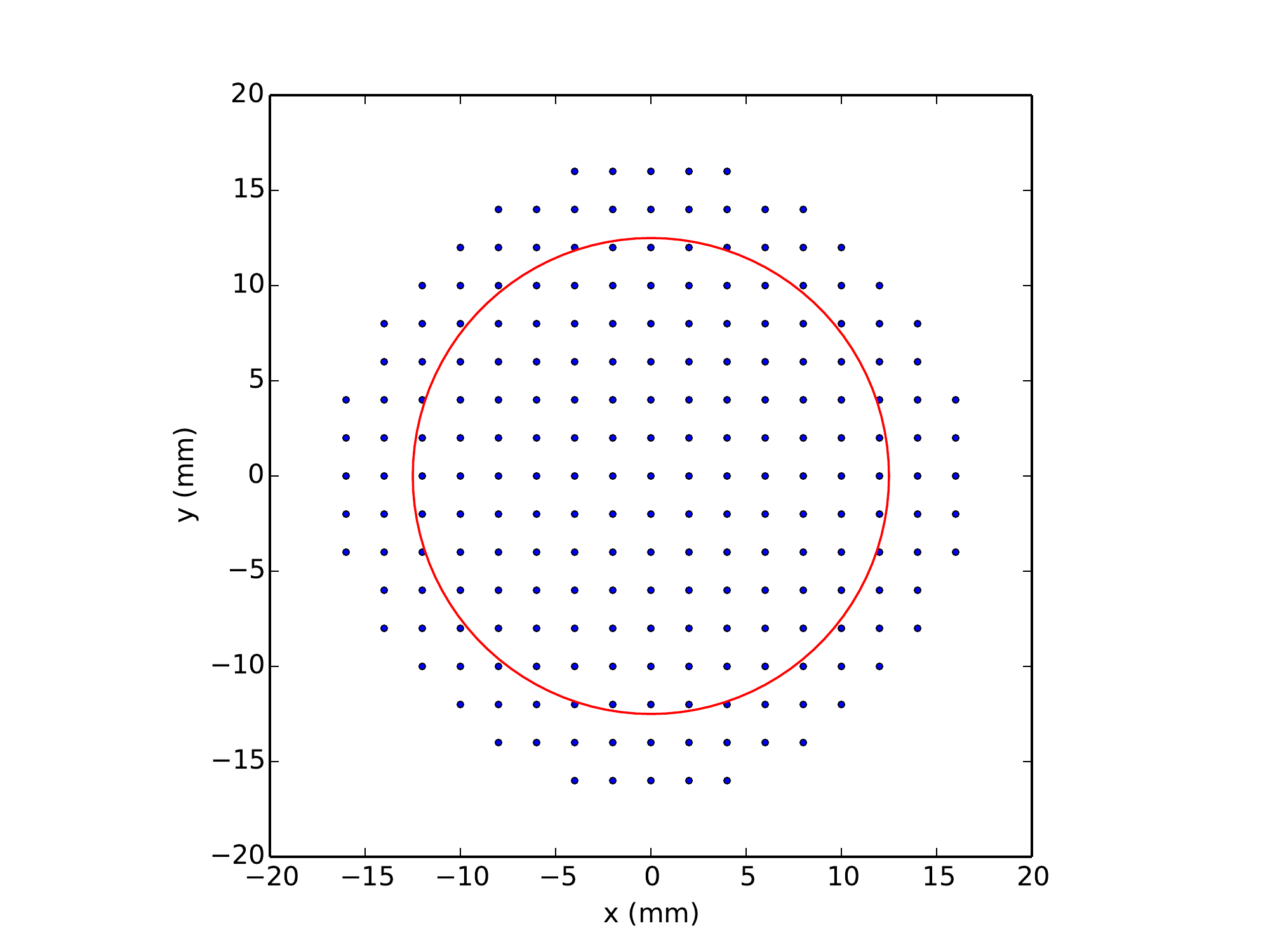}}
\vspace{0cm}
\caption{
Coordinates for all points covered in a complete scan. 
At each point a run of 8000 flasher events is taken.
The red circle indicates the nominal size (25 mm diameter) 
and position of the photocathode of a PMT under test.
}
\label{scan_points}
\end{figure}

After this, a 221-point scan with 2 mm pitch was carried out.
The scan points are shown in Fig.~\ref{scan_points}.

\section{Data Analysis}

\begin{figure}[h!]
\centerline{\includegraphics[width=1.0\linewidth]{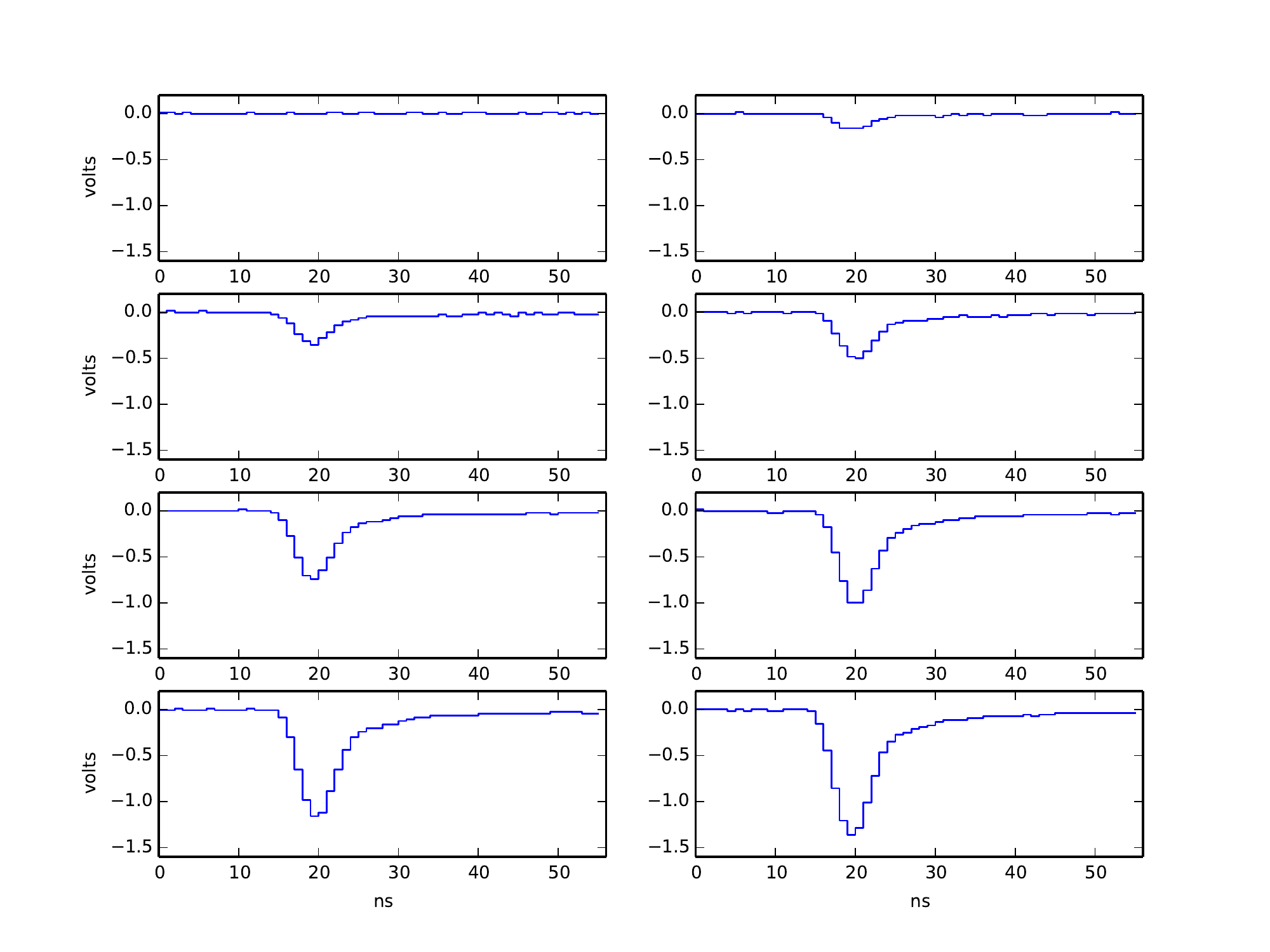}}
\vspace{0cm}
\caption{
Traces for a cycle of 8 flasher events for the monitor PMT, showing
the pulse shapes for different levels of illumination, starting with 
no LEDs firing and finishing with 7 LEDs firing.
Each trace comprises 56 samples, taken at 1 ns intervals. 
The average of the first 10 samples is used as a pedestal and has been 
subtracted from all the samples.
The magnitude of the sum of all pedestal-subtracted samples is used as the 
charge of the pulse.
}
\label{traces}
\end{figure}

\begin{figure}[h!]
\vspace{0cm}
\centerline{\includegraphics[width=1.1\linewidth]{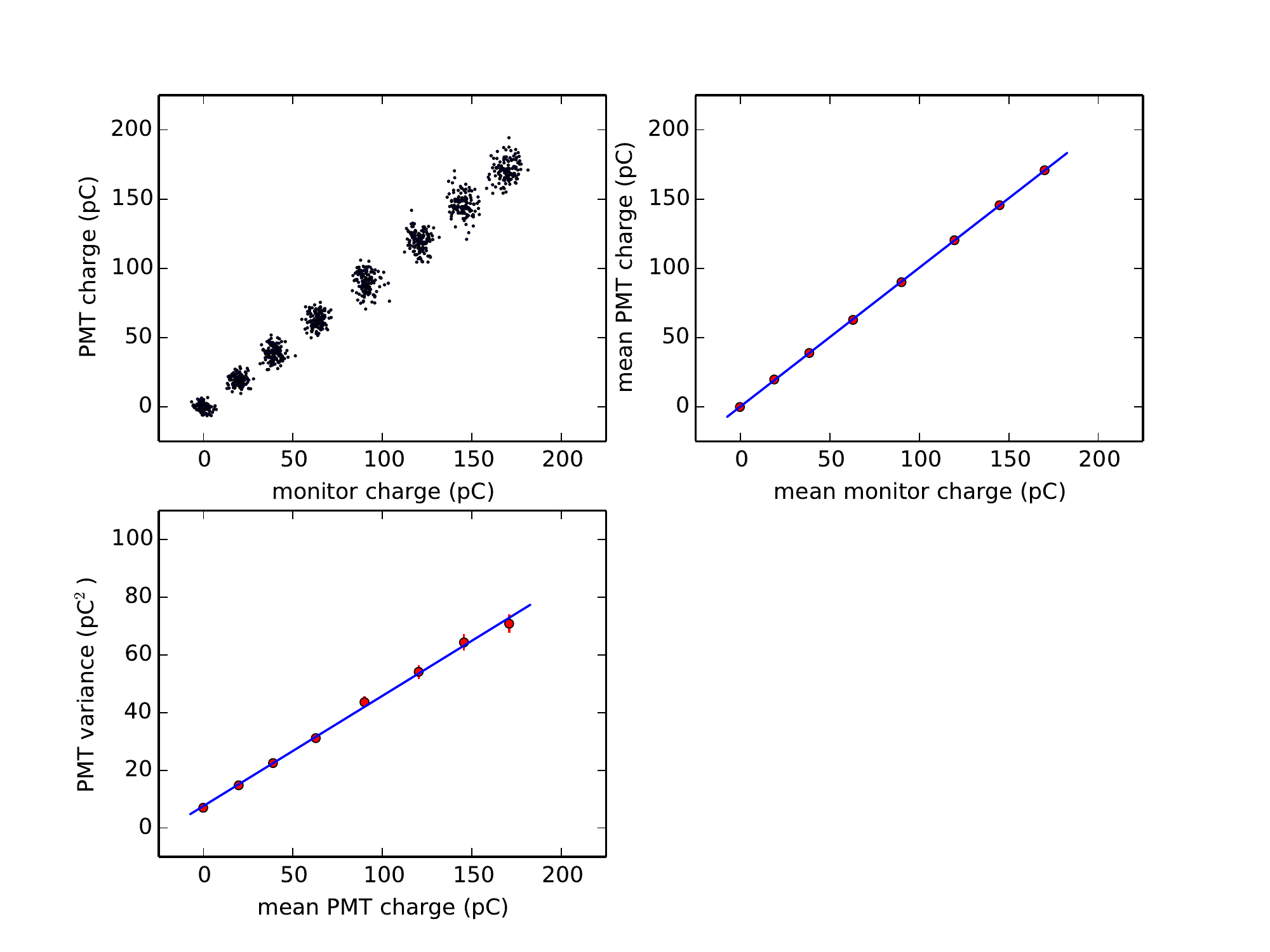}}
\vspace{0cm}
\caption{
Sample data for an arbitrary test PMT at a single scan point.
Upper left: the charges for the test PMT are plotted vs those from the 
monitor PMT.
Upper right: mean values  from the test PMT distributions vs mean 
values from the monitor PMT distributions.
The slope of the fitted line is used as an estimator of
the relative gain of the PMT.
Lower left: $\sigma^2$ values of the test PMT charge distributions vs 
the mean values of those distributions.
The slope of the fitted line is used as an estimator of the photostatistics
gain of the PMT.
}
\label{sample_1}
\end{figure}

At each scan point, and for each PMT, there are 8000 flasher events, 1000 
at each level of illumination.
One can see this in the upper-left plot in Fig.~\ref{sample_1} where the 
charge for an arbitrary test PMT is plotted against the charge from the same
event coming from the monitor PMT. 
The charges are integrals of the pedestal-subtracted traces from the digitizers.
Example traces are shown in Fig.~\ref{traces}. 

After the charges from the PMTs for all the events are computed, the relative 
gains and photostatistics gains are calculated.
The relative gain includes all components in the PMT and readout path, 
including the photocathode, and is defined 
as the slope of the test PMT mean charge vs monitor PMT mean charge.
An example is shown in the upper-right plot in Fig.~\ref{sample_1} where each 
cluster of points seen in the upper-left plot is reduced to a mean value and
uncertainty. 

The photostatistics gain depends on components 
downstream from the photocathode, the
most important (and variable) being the amplification due to the dynodes.
The simple model for determining this gain 
relies on the assumption that the width of
the charge distribution, for a constant illumination level, is caused by 
Poisson fluctutations in photoelectron production. 

Let $N_{pe}$ be the mean number of photoelectrons produced. Poisson 
fluctuations will produce an approximately Gaussian distribution 
with $\sigma = \sqrt{N_{pe}}$. 
The dynodes and any further downstream amplification 
(such as the preamp integrated into the
PMT assembly) will multiply the
charges by a factor $G$.
This results in a charge distribution with mean = $\mu$ and RMS = $\sigma$ where
$\mu = G N_{pe}$ and $\sigma = G \sqrt{N_{pe}}$. 
Thus the slope of $\sigma^2$ plotted vs $\mu$ will be $G$.
 
This is a simplification; noise in the electronics will contribute 
(in quadrature) to $\sigma$, as will any fluctuations in the light source.
There is also an effect arising from fluctuations in the gain for a single 
electron (single electron response or $SER$ in ~\cite{bencheikh}). 
A correction factor, $1 + \delta_{SER}^2$ where $\delta_{SER} = 
\sigma_{SER}/\mu_{SER}$ with $\mu_{SER}$ the mean of the 
single-photoelectron distribution and $\sigma_{SER}$ its width,
is needed for a more exact estimate of the absolute gain.
Single-photoelectron measurements are made routinely with the VERITAS flasher
systems ~\cite{flasher} and indicate that the PMTs have an average width-to-mean ratio of 
about 0.35 which means a correction of 10-15\% is needed. 
Single-photoelectron measurements were not made in the study reported here and
the correction is not applied.

The expression used here is a convenient 
approximation for our purposes, which are 
concerned more with the relative behaviour of two groups of PMTs. 
A sample $\sigma^2$ vs $\mu$ plot is shown in the lower left of 
Fig.~\ref{sample_1}.
The slope of the fit is $0.39 \pm 0.02$ pC/photoelectron.
Accounting for the preamp gain (approximately 6.6) we have $(3.7 \pm 0.2) \times 10^5$ 
for the gain supplied by the dynodes.

\subsection{Photocathode Studies}

\subsubsection{Uniformity}

\begin{figure}[h!]
\centerline{\includegraphics[width=1.0\linewidth]{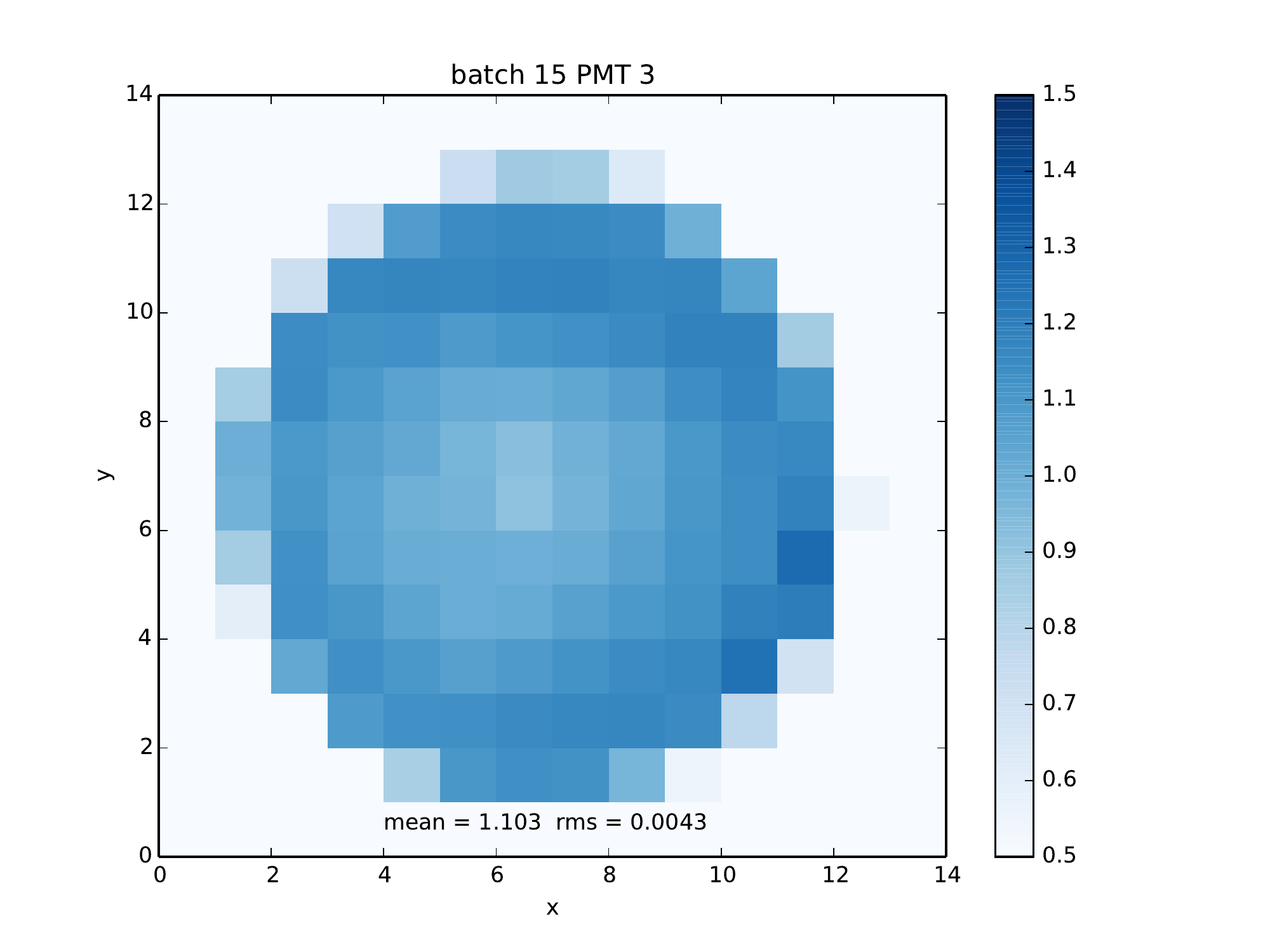}}
\vspace{0cm}
\caption{
A plot of the relative gain, as defined in Fig.~\ref{sample_1}, 
at each scan point for a typical PMT.
The mean and RMS for the highest 90 values for the PMT
are displayed.
The RMS is a measure of the uniformity of the photocathode.
}
\label{scan-15-3-a}
\end{figure}

\begin{figure}[h!]
\centerline{\includegraphics[width=1.2\linewidth]{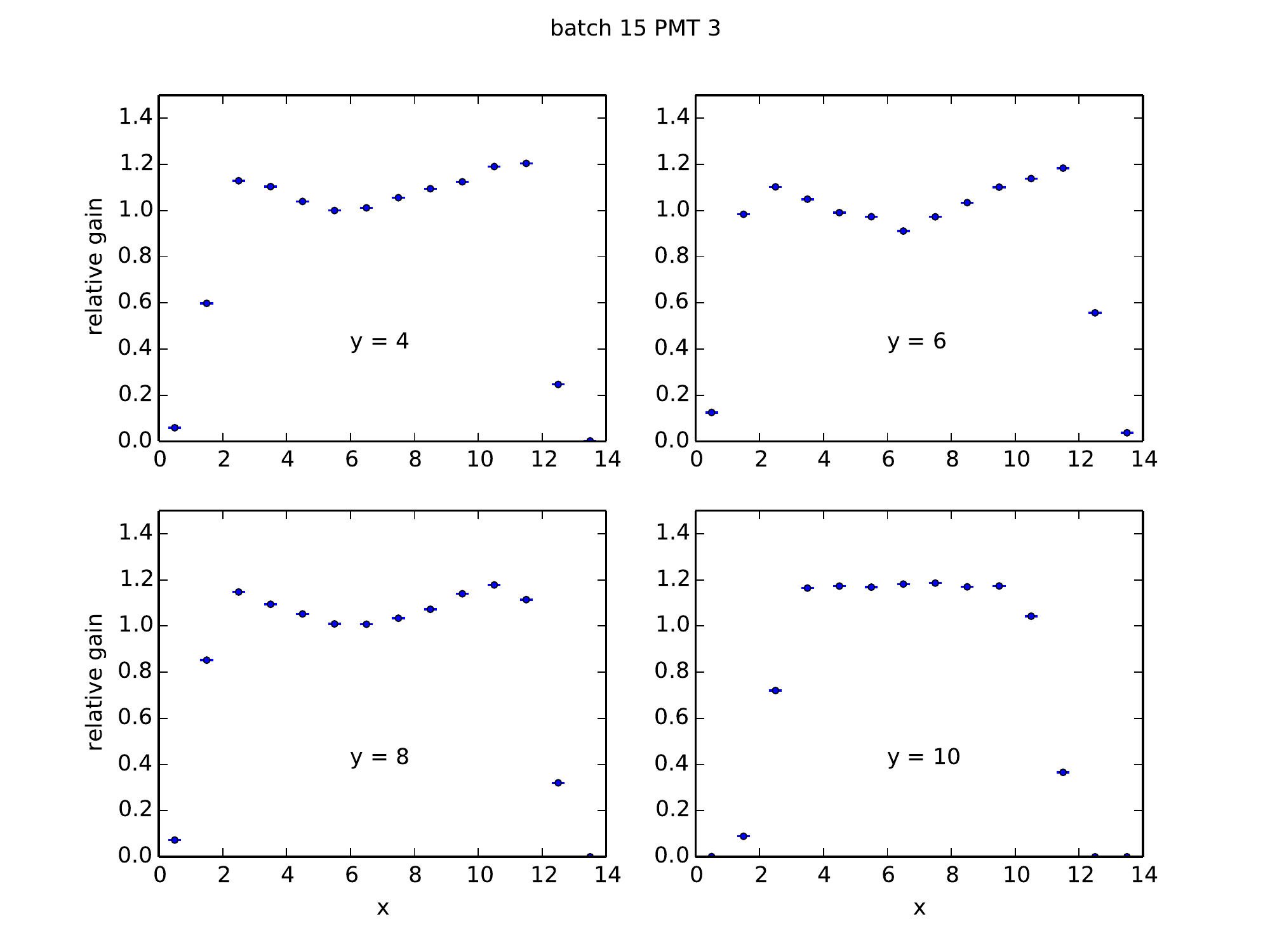}}
\vspace{0cm}
\caption{
Sample rows for the data shown in Fig.~\ref{scan-15-3-a}.
Statistical uncertainties on the relative gains are smaller 
than the symbol size. 
}
\label{scan-15-3-b}
\end{figure}

The relative gain can be computed for every scan point on every PMT. 
The resulting numbers can then be displayed, as in Figs.~\ref{scan-15-3-a}
and~\ref{scan-15-3-b}.
Changes from place to place on the photocathode are smooth and slowly varying.
There is no evidence for large holes or other signs of damage. 

\begin{figure}[h!]
\centerline{\includegraphics[width=1.2\linewidth]{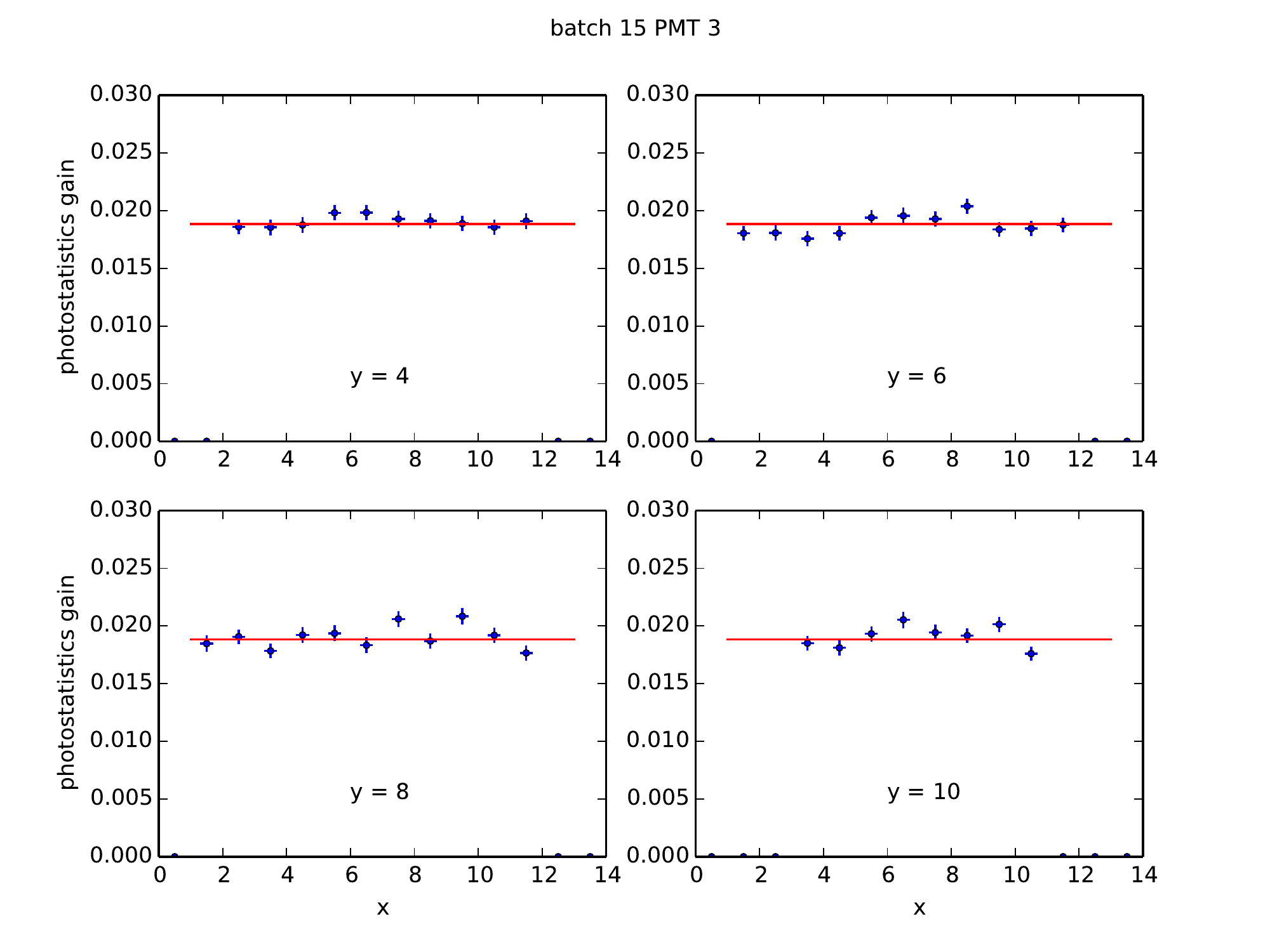}}
\vspace{0cm}
\caption{
Photostatistics gains for the same scan points as used in 
Fig.~\ref{scan-15-3-b}.
The horizontal line indicates the average of photostatistics 
gains from all points
within the lateral limits of the photocathode.
All points are consistent with the global average.
}
\label{scan-15-3-c}
\end{figure}

Another way to look at the uniformity is to show the relative gains 
in a series 
of plots, each one displaying a row in x at constant y.
An example of this is shown in Fig.~\ref{scan-15-3-b}.
A significant amount of variation is seen.
This is presumably due to nonuniformities in the photocathode and 
position-dependent focussing from the cathode to the first dynode.

One can do this for the photostatistics gains also, as in 
Fig.~\ref{scan-15-3-c}.
The global average of the photostatistics gains determined at each scan point
within the lateral limits of the photocathode 
is plotted as a horizontal line in each subplot.
All the points are consistent with the average value.
This indicates that the photostatistics gains depend only on the dynodes 
(and components further downstream) and are independent of where on 
the photocathode the light is being injected.

\begin{figure}[h!]
\vspace{0.0cm}
\centerline{\includegraphics[width=1.1\linewidth]{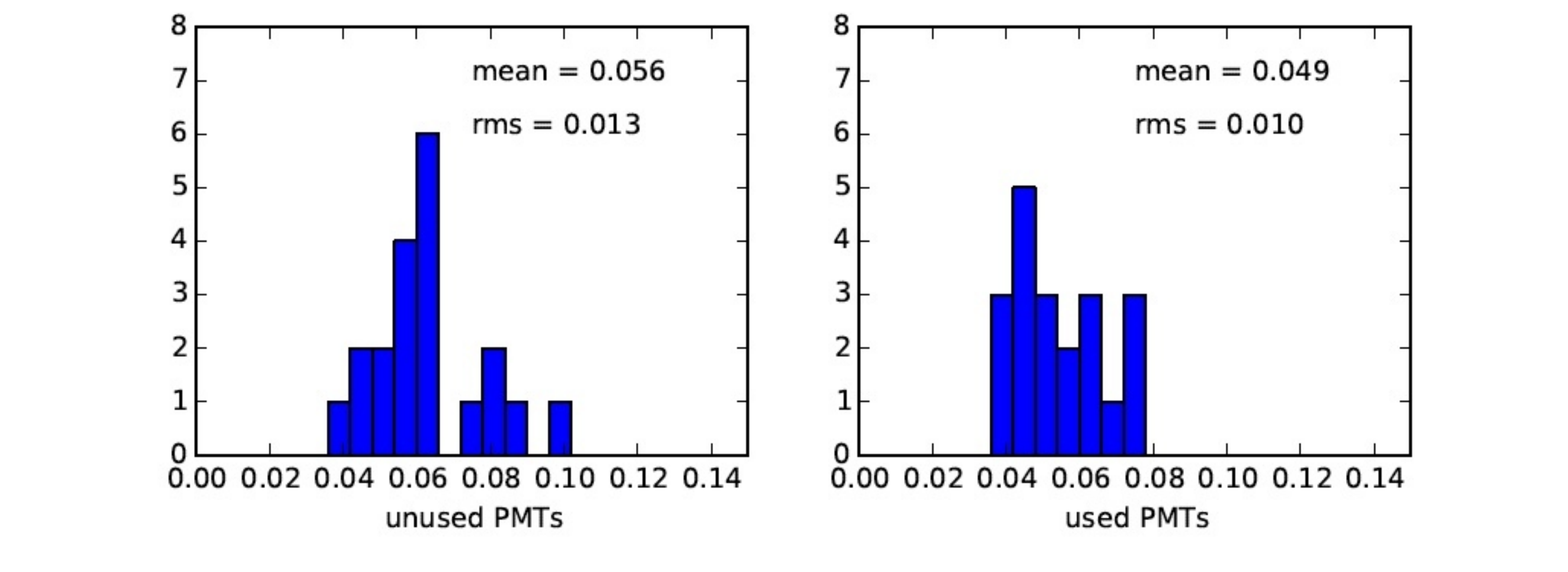}}
\vspace{0.0cm}
\caption{
The RMS values of the relative gains, over the PMT faces, from data like
that shown in Fig.~\ref{scan-15-3-a}.
The distribution for the 20 unused PMTs is shown on the left 
and that for the 20 used PMTs is shown on the right. 
The means and RMS values for the two distributions not significantly 
different.
}
\label{rms_new_old}
\end{figure}

A convenient parameter for estimating the uniformity of 
the photocathode is the RMS of the relative gains corresponding 
to scan points that are within the boundaries of the photocathode. 
This quantity is plotted in Fig.~\ref{rms_new_old}. 

\subsubsection{Relative Quantum Efficiency}

Another diagnostic is the ratio of the relative gain to the 
photostatistics gain.
As mentioned earlier, the relative gain includes effects of the photocathode
but the statistics gain does not.
The ratio thus contains information about the photocathode.

Referring to Fig.~\ref{sample_1}, the slope of the line shown in the lower
left panel is $G$, the photostatistics gain.
The slope of the line in the upper right panel is the relative gain,
$G \epsilon_q k$,
where $\epsilon_q$ is the quantum efficiency of the part of the photocathode
illuminated for the scan point corresponding to the data displayed and 
$k$ is a factor containing all the details relating the size of the monitor 
signal to the number of photons hitting the photocathode.
We assume that $k$ is constant but do not know its value and therefore cannot
measure the absolute value of $\epsilon_q$.

\begin{figure}[h!]
\vspace{0.0cm}
\centerline{\includegraphics[width=1.1\linewidth]{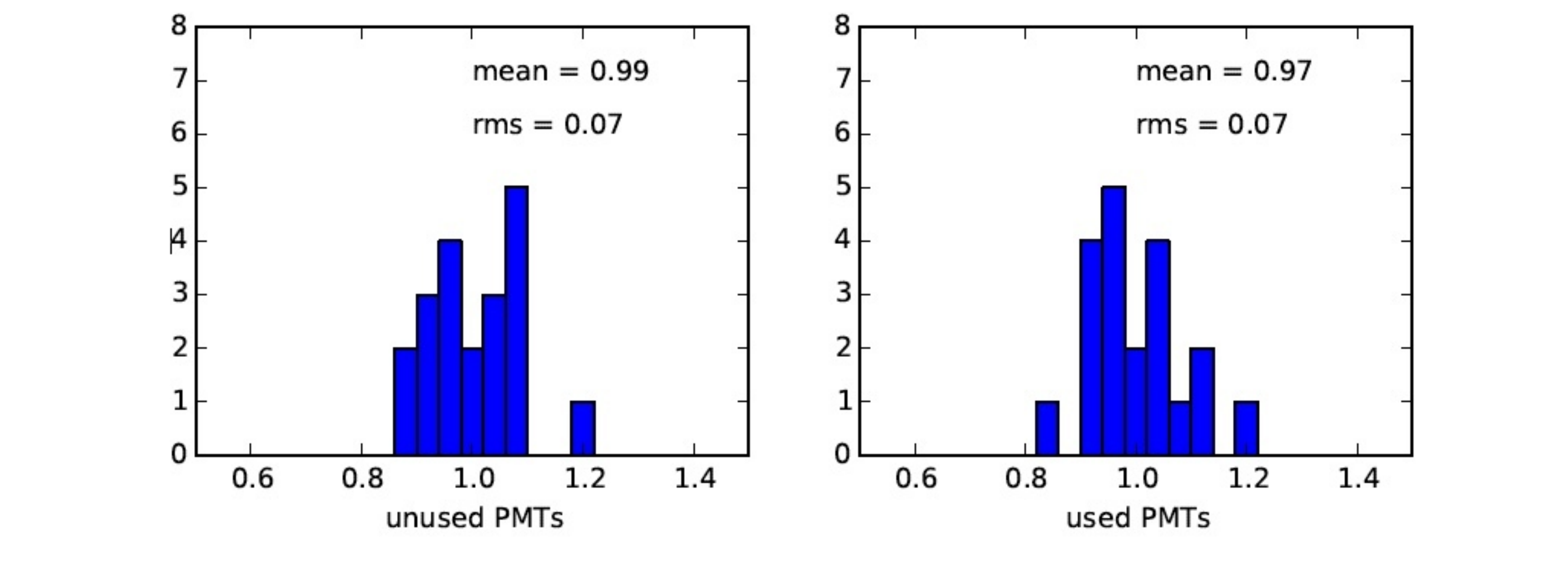}}
\vspace{0.0cm}
\caption{
As in Fig.~\ref{rms_new_old} but for the relative quantum efficiencies
as described in the text.
}
\label{rel_qe}
\end{figure}

To obtain the relative value of $\epsilon_q$ for each PMT we first compute the 
ratio of the average relative gain of that PMT to the average 
photostatistics gain of the PMT.
We scale the ratios for the 40 PMTs to have an average of unity
to factor out our ignorance of $k$.
The results are plotted in  Fig.~\ref{rel_qe}. 
Here, as in Fig.~\ref{rms_new_old}, 
one sees no significant difference between the used and unused PMTs.

\subsection{Dynode Studies}

\begin{figure}[h!]
\vspace{0.0cm}
\centerline{\includegraphics[width=1.1\linewidth]{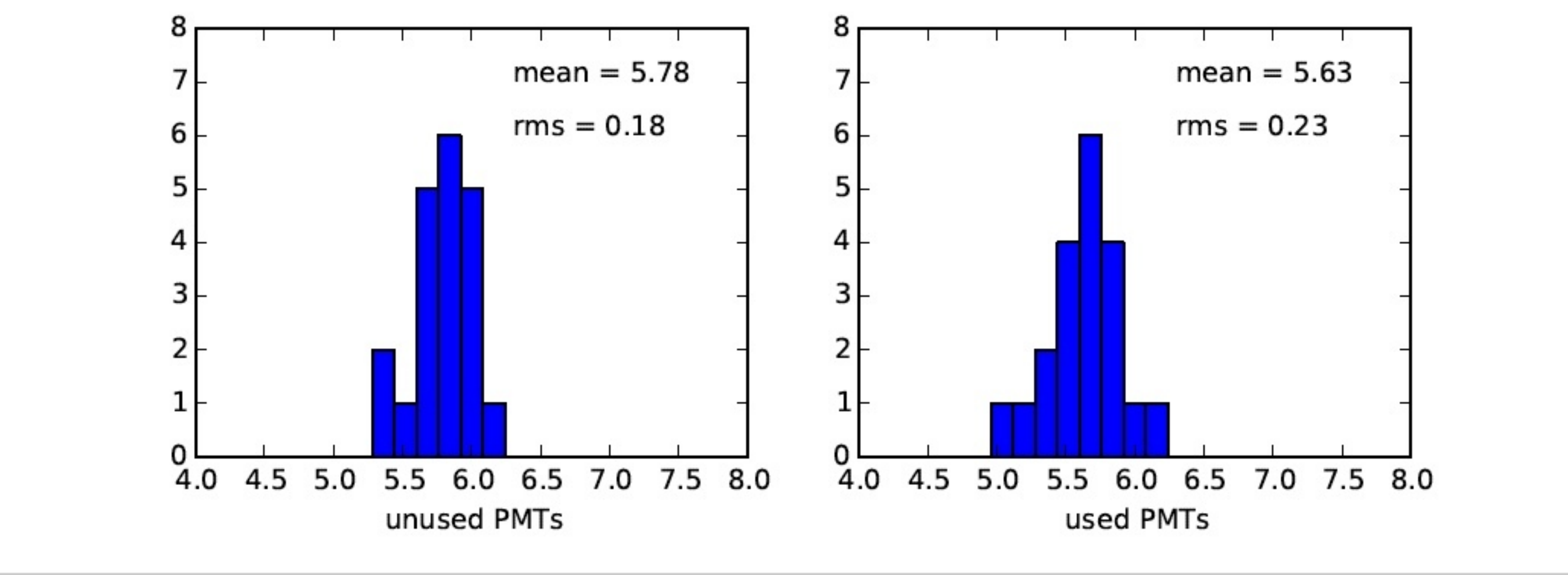}}
\vspace{0.0cm}
\caption{
As in Fig.~\ref{rms_new_old} but for the $\alpha$ parameter used in 
parameterizing the high-voltage dependence of the gains.
}
\label{power_new_old}
\end{figure}

As explained earlier, we used the function $G/G_0 = (HV/HV_0)^\alpha$
to parameterize the HV dependence of the PMT gains.
The $\alpha$ parameter is related to the number of dynodes (8 in the 
PMTs studied here) and their secondary emission characteristics so it 
could contain information about their state; an anomalous value 
would arise if one or more steps in the amplification
chain were compromised. 
This argument motivates Fig.~\ref{power_new_old} where the $\alpha$ values
for all of the tested PMTs are plotted, with those from the unused PMTs on the 
left and from the used PMTs on the right. 

As with the relative-gain RMS values shown in Fig.~\ref{rms_new_old}, there is 
no systematic difference between the two sets. 
This supports the notion that any dynode damage due to prolonged use is slight 
and can be mitigated by increasing the high voltage on a yearly basis.

\section{Conclusions}

Two groups of 20 PMTs have been tested using a flasher system and a device which
enables one to scan over the surface of six PMTs in parallel.
One group comprised PMTs that have been in use at VERITAS since 2012 while 
the other group was made up of PMTs of the same type 
that have been kept as spares and never used.
In terms of gains, photocathode efficiency and uniformity, and variation with
high-voltage settings, both groups of PMTs are very similar.
As seen in Figs.~\ref{rms_new_old}, ~\ref{rel_qe}, and ~\ref{power_new_old},
there is significant scatter of these measures, within the two
groups of PMTs. 
Their mean values are slightly different, but the
differences are consistent with the statistical fluctuations expected from the
relatively small sample size used in this study.

We conclude that the PMTs used in the VERITAS telescope cameras have not
been adversely affected over the years and can be expected to continue operating
well for a similar or longer period in the future.

\section{Acknowledgements}

VERITAS research is supported by grants from the U.S. DoE, the NSF, and the 
Smithsonian, by NSERC in Canada, and by the Helmholtz Association in Germany. 
We thank S. Kumar for help during the initial stages of this work and our 
colleagues in the VERITAS collaboration
who extracted the used PMTs from the cameras under
difficult conditions brought on by the COVID-19 pandemic.

\end{document}